\documentstyle[a4,11pt,epsf,draft]{article}

\begin{document}

\title{High energy description of dark energy in an approximate 3-brane Brans-Dicke
cosmology}
\author{A.
Errahmani{\footnote{a.errahmani@sciences.univ-oujda.ac.ma}},$\;\;\;$
 T. Ouali{\footnote{ouali@sciences.univ-oujda.ac.ma}}\\
Department of Physics, Faculty of Sciences, University Mohammed I,
\\BP 717, Oujda, Morocco} \maketitle

\begin{abstract}
We consider a Brans-Dicke cosmology in five-dimensional
space-time. Neglecting the quadratic and the mixed Brans-Dicke
terms in the Einstein equation, we derive a modified wave equation
of the Brans-Dicke field. We show that, at high energy limit, the
3-brane Brans-Dicke cosmology could be described as the standard
one by changing the equation of state. Finally as an illustration
of the purpose, we show that the dark energy component of the
universe agrees with the observations data.

\end{abstract}

\section{\bf{Introduction:}}\par
Theories, in which scalar fields are coupled directly to the
curvature, are termed scalar-tensor gravity, such as the
low-energy effective string theory \cite{Green} which couples a
dilaton field
 to the Ricci curvature tensor. However, the simplest and the best known is the
Brans-Dicke (BD) theory \cite{Brans}. The BD theory, which is a
generalization of the general relativity, must recover the latter
as the BD parameter $\omega$ goes to infinity \cite{Wein}. From
timing experiments using the Viking space probe \cite{Rea},
$\omega$ must exceed 500. This constraint ruled out many of
extended inflation \cite{EWein,La} and provides a succession of
improved models of extended inflation
\cite{Hol1,Hol2,Bar,Stein}.\par
 Superstring theory suggests that
the space-time of our universe might be of higher dimension
\cite{H1,H2}, in which the extra dimensions are compactified and
only the 4-dimensional (4D) is observed experimentally. Recently,
a great deal of interest has been done in cosmological scenario in
which matter field are confined to a 3-brane world imbedded in a
5-dimensional (5D) bulk space-time \cite{R1,R2,Sch}.\par
 Dark energy, distributed homogeneously in
the Universe, is a component of the critical density of our
universe as showed by the cosmic microwave background (CMB) and
type Ia supernovae (SNe) observations \cite{Sp,Perl,Kn}. Using
type Ia SNe \cite{Perl} as standard candles to gauge the expansion
of the universe shows that the dark energy causes the expansion of
the universe to speed up. This two experiments are in agreement
with $\Omega_\Lambda\simeq 0.7$ and $\Omega_m\simeq 0.3$ .

In this paper, the dynamical system and dark energy of the
universe are studied using the BD field in the 3-brane world. In
brane cosmology \cite{R1,R2,Sch}, at the very early time, i.e. at
high energy limit, dynamical evolution of the universe is modified
by the extra terms in the Einstein equations, otherwise by the
square of the matter density on the brane. In this way, we
generalize the 4D BD theory to the 5D one by considering that BD
field is sensitive only to the physical 3-brane. So it is
described by the same 4D action  and must recover the standard BD
cosmology at low energy. To this aim, we add simply a BD stress
energy tensor to the modified Einstein equations, $^4G_{\mu\nu}$,
in 5D bulk space-time by neglecting all quadratic and mixed terms
of this stress-energy tensor. To illustrate our consideration, we
follow first Kolitch's work \cite{K2} in order to show, first,
that at high energy limit the 5D space-time could be described by
4D space-time cosmology, i.e. the information contained in the
extra dimensions are now involved in the equation of state. The
$\gamma$ factor characterizing the matter content of the universe
in standard BD cosmology is equal to twice of the one in the
3-brane. Second, following \c{C}alik's work \cite{C1}, we show the
contribution of the dark
energy in the dynamical evolution of the universe.\\
The paper is organized as follows: a short review of the standard
BD cosmology with vacuum cosmological constant as well as the
3-brane world with BD field is presented in section 2. Section 3
is devoted to resolve the dynamical system of the universe, while
in section 4 we relate the cosmological parameters to the dark
energy. A conclusion is given in section 5.

\section{ \bf BD cosmology with cosmological constant:}
\subsection{\bf 4-dimensional BD cosmology:}

Brans-Dicke cosmology with a nonzero cosmological constant,
$\Lambda_4$, was studied by many authors \cite{K2,K1,W,Q}. In this
section, we follow the notations and the work of the author
\cite{K2} where the action has the form:

\begin{equation}\label{action}
S=\int dx^{4}\sqrt{-g}\left( \phi \left[ R-2\Lambda_4 \right]
-\frac{\omega }{\phi }\partial _{\mu }\phi \partial ^{\mu }\phi
-16\pi {\mathcal{L}}_{m}\right)
\end{equation}

By varying  this action with respect to the metric and BD field
$\phi$, the homogeneous and isotropic Friedmann-Robertson-Walker
equation with scale factor a(t) and spatial curvature index $k$
and the wave equation of the BD field are:

\begin{equation}\label{FRW1}
(\frac{\dot{a}}{a}+\frac{\dot{\phi }}{2\phi
})^{2}+%
\frac{k}{a^{2}}=\frac{3+2\omega }{12}\left( \frac{\dot{\phi }%
}{\phi }\right) ^{2}+\frac{8\pi \rho }{3\phi }+\frac{\Lambda_4
}{3}
\end{equation}

and

\begin{equation}\label{FRW2}
-\frac{1}{a^{3}}\frac{d(\dot{\phi
}a^{3})}{dt}=\frac{8\pi }{%
3+2\omega}((3\gamma-4)\rho-\frac{\Lambda_4 \phi }{4\pi })
\end{equation}
where $\omega$ is the BD parameter, the dot denotes the derivative
with respect to the time and $T_{\mu \nu }={\rm{diag}}(\rho
,p,p,p)$ is the stress-energy tensor of a perfect fluid in an
orthonormal frame. From the conservation equation of $T_{\mu \nu
}$ we have:

\begin{equation}\label{FRW3}
\dot{\rho }=-3\frac{\dot{a}}{a}(\rho +p)
\end{equation}

Equations (\ref{FRW1}), (\ref{FRW2}) and (\ref{FRW3}) could be
rewritten, as in \cite{K2,K1}:

\begin{equation}\label{X1}
\dot{X}=3\frac{Y^{2}}{A}(\gamma /2-1)-\frac{3\gamma
X^{2}}{2}+%
\frac{XY}{2A}+\frac{\Lambda_4 \gamma }{2}
\end{equation}

\begin{equation}\label{Y1}
\dot{Y}=X^{2}(1-3\gamma /4)-\frac{Y^{2}}{2A}(1-3\gamma
/2)-3XY-%
\frac{\Lambda _4}{6}(1-3\gamma /2)
\end{equation}

where the new variables are defined as:

\begin{equation}
X=\left( \frac{\dot{a}}{a}+\frac{\dot{\phi
}}{%
2\phi }\right)
\end{equation}

\begin{equation}
Y=A\frac{\dot{\phi }}{\phi }; \qquad\qquad{\rm{and}} \ A=\left(
3+2\omega\right) /12.
\end{equation}

The equation of state is given by

\begin{equation}\label{state}
p=(\gamma -1)\rho
\end{equation}

The solutions of this planar dynamical system have previously been
examined in \cite{K2,C1,K1} \par
 In the next subsection we will be
interested in this dynamical system but in an extra dimension
particularly in the 3-brane world.

\subsection{\bf 5-dimensional BD
 cosmology:}

The modified Einstein equations on the 3-brane, derived from
5-dimensional bulk space-time, have the form \cite{Sch}
\begin{equation}\label{5ee}
    ^4G_{\mu\nu}=-\Lambda_4q_{\mu\nu}+k_4^2\tau_{\mu\nu}+k_5^4\Pi_{\mu\nu}
    -E_{\mu\nu}
\end{equation}
where $\Lambda_4
=\frac{1}{2}k_5^2(\Lambda_5+\frac{1}{6}k_5^2\lambda^2)$ is the 4D
cosmological constant, $k_4^2=8\pi G_N=\frac{k_5^4\lambda}{6\pi}$
and $k_5^2$ are the 4D and the 5D gravitational constant
respectively ($G_N$ is the Newton's constant of gravity) and the
quadratic tensor $\Pi_{\mu\nu}$ is given by:

\begin{equation}\label{}
\Pi_{\mu\nu}=-\frac{1}{4}\tau_{\mu\alpha}\tau^\alpha_{\nu}+
\frac{1}{12}\tau\tau_{\mu\nu}+
\frac{1}{8}q_{\mu\nu}\tau_{\alpha\beta}\tau^{\alpha\beta}
-\frac{1}{24}q_{\mu\nu}\tau^2
\end{equation}
$E_{\mu\nu}$ is a part of the 5D Weyl tensor which we take equal
to zero. $\Lambda_5$ is the cosmological constant of the bulk
space-time. $\lambda$, $\tau_{\mu\nu}$ and $q_{\mu\nu}$ are the
tension, the energy momentum tensor and the
metric, respectively, confined on the brane world.\\
 Our proposal to generalize the gravitational equation
 (\ref{5ee}) in 3-brane world is done as follows. First, to obtain the
 equivalent 4D BD of the field equations
 in the 5-dimensions universe, we consider
that the behaviour of BD field is sensitive only to physical
3-brane, so it is described by the same action as in 4-dimension,
Eq. (\ref{action}). Second, to recover the BD cosmology at low
energy, we propose to add simply a BD stress energy tensor to the
Einstein equation (\ref{5ee}) in 5D bulk space-time, where all
quadratic and mixed terms of this stress-energy tensor are
neglected. So the modified Einstein equations are then written as:

\begin{eqnarray}
    ^4G_{\mu\nu}&=&-\Lambda_4q_{\mu\nu}+8\pi G_N\tau_{\mu\nu}+k_5^4\Pi_{\mu\nu}
    -E_{\mu\nu}\\
 \nonumber    &+&\frac{\omega}{\phi^2}(\phi;_\mu\phi;_\nu-\frac{1}{2}q_{\mu\nu}
    \phi;_\lambda\phi^{;\lambda})+\frac{1}{\phi}(\phi;_{\mu;\nu}-q_{\mu\nu}\Box\phi)
\end{eqnarray}
The BD field equations obtained by varying the action $S$, Eq.
(\ref{action}), with $E_{\mu \nu }=0$ and $G_N=\frac{1}{\phi}$,
are:

\begin{equation}
(\frac{\dot{a}}{a}+\frac{1}{2}\frac{\dot{\phi }}{%
\phi })^{2}+\frac{k}{a^{2}}=\frac{3+2\omega }{12}\left(
\frac{%
\dot{\phi }}{\phi }\right) ^{2}+\frac{8\pi \rho }{3\phi }+\frac{%
k_{5}^{4}}{36}\rho ^{2}+\frac{\Lambda _{4}}{3}
\end{equation}

\begin{equation}\label{mwe}
-\frac{1}{a^{3}}\frac{d(\dot{\phi
}a^{3})}{dt}=\frac{8\pi }{%
3+2\omega }((3\gamma -4)\rho +\frac{k_{_{5}}^{^{4}}\phi }{48\pi
}\left( 3\gamma -2\right) \rho ^{2}-\frac{\Lambda _{4}\phi }{4\pi
})
\end{equation}

\begin{equation}
\dot{\rho }=-3\frac{\dot{a}}{a}\gamma \rho
\end{equation}
The wave equation of the BD field, (\ref{mwe}), in the 3-brane
world differs from the one, Eq. (\ref{FRW2}), in the standard BD
cosmology. This means that the extra dimensions affect not only
the Einstein equations but also the wave equation and the equation
of state as will be
stated later. \\
So at low energy limit, ${\protect\rho \gg \protect\rho }^{2}$,
 the field equations are the same as
(\ref{FRW1}) and (\ref{FRW2}); or equivalently in terms of the
variables X and Y, we recover the equations (\ref{Y1}) and
(\ref{X1}).

At high energy limit,  ${\protect\rho \ll \protect\rho }^{2}$, the
field equations become:

\begin{equation}
(\frac{\dot{a}}{a}+\frac{1}{2}\frac{\dot{\phi }}{%
\phi })^{2}+\frac{k}{a^{2}}=\frac{3+2\omega }{12}\left(
\frac{%
\dot{\phi }}{\phi }\right) ^{2}+\frac{k_{5}^{4}}{36}\rho
^{2}+\frac{%
\Lambda _{4}}{3}
\end{equation}

\begin{equation}
-\frac{1}{a^{3}}\frac{d(\dot{\phi
}a^{3})}{dt}=\frac{8\pi }{%
3+2\omega }(\frac{k_{_{5}}^{^{4}}\phi }{48\pi }\left( 3\gamma
-2\right) \rho ^{2}-\frac{\Lambda _{4}\phi }{4\pi })
\end{equation}

\begin{equation}
\dot{\rho }=-3\frac{\dot{a}}{a}\gamma \rho
\end{equation}

In term of the variables X and Y, we recover the following
dynamical system:

\begin{equation}
\dot{X}=3\frac{Y^{2}}{A}(\gamma -1)-3\gamma
X^{^{2}}+\frac{XY}{%
2A}+\Lambda _{4}\gamma
\end{equation}

\begin{equation}
\dot{Y}=X^{^{2}}\left( 1-\frac{3}{2}\gamma \right)
-\frac{%
Y^{^{2}}}{2A}\left( 1-3\gamma \right) -3XY-\frac{\Lambda
_{4}}{6}\left( 1-3\gamma \right)
\end{equation}

 which are the same as (\ref{Y1}) and (\ref{X1}) with $\gamma$ in the standard
 case is equal to twice of the one in 5D bulk space-time. Therefore the 3-brane BD
 theory at high energies limit, could be described
by the 4D one with the following equation of state
\begin{equation}\label{}
    p=(2\gamma-1)\rho
\end{equation}

\bigskip

\section{Dynamical system of the universe:}

 To show how the dark energy contributes to the dynamical system of the
 universe, and how changes appear from 4D to 5D; we linearize the dynamical
 system about the stable cosmological non vacuum solution with flat space and show
how the Hubble parameter varies with the scale factor a(t).

\subsection{ \bf  Equilibrium solutions
for a flat space:}

To study the dark energy in four and five dimensional BD
cosmology using the Hubble parameter, we follow the work of \cite{C1} by introducing the variables, $H=%
\frac{\dot{a}}{a}$ and $F=\frac{\dot{\phi }}{\phi }$ rather than
$X$ and $Y$. The field equations (\ref{FRW1}) and (\ref{FRW2})
become:

\begin{eqnarray}\label{HF1}
\nonumber aH\left( 2\omega +3\right) H^{\prime } &=&-3\left(
\gamma \omega +2\right) H^{2}-\omega \left( \frac{\omega \left(
2-\gamma \right) +1}{2}\right)
F^{2}-\omega (3\gamma -4)HF  \\
 &&-\frac{k}{a^{2}}\left( \omega \left( 3\gamma -2\right)
+3\right) +\Lambda _{4}\left( \gamma \omega +1\right)
\end{eqnarray}

\begin{eqnarray}\label{HF2}
\nonumber aH\left( 2\omega +3\right) F^{\prime } &=&-3\left(
3\gamma -4\right) H^{2}-\left( 4\omega +3-\frac{3\omega \gamma
}{2}\right) F^{2}-3(3\gamma
-1+2\omega )HF   \\
 &&-\frac{3k}{a^{2}}\left( 3\gamma -4\right) +\Lambda
_{4}\left( 3\gamma -2\right)
\end{eqnarray}\\

where the prime denotes the derivative with respect to the scale factor. \\
 Since the term $k/a^{2}$ decreases dramatically as $a(t)$
increases with the expansion of the universe, we drop it in the
next by taking $k=0$.

Let $(H_{\infty },F_{\infty })$ be the equilibrium points for a
flat space $\left( k=0\right)$, which are obtained by setting H'
and F' equal to zero in equations (\ref{HF1}) and (\ref{HF2}). The
solutions of the resulting equations are:

\begin{equation}
\left( H_{\infty }\left[ 1\right] ,F_{\infty }\left[ 1\right]
\right)
=\sqrt{%
\frac{6\Lambda_4 }{4+6\omega \gamma -3\omega \gamma ^{2}}}\left(
-1/3,\gamma \right)
\end{equation}

and

\begin{equation}
\left( H_{\infty }\left[ 2\right] ,F_{\infty }\left[ 2\right]
\right)
=\sqrt{%
\frac{2\Lambda_4 }{\left( 2\omega +3\right) \left( 3\omega
+4\right) }}\left( \omega +1,1\right) \allowbreak
\end{equation}

the $\left( \infty \right) $ index means that the equilibrium is
taken for a late time expansion.

\bigskip The first solution is unstable while
the second one is stable. In the next we will be interested only
in the second solution and we must have $\omega > -4/3$ or $\omega
< -3/2$, and in the limit $\omega \longrightarrow +\infty $ $\ $
we have:

\begin{equation}\label{Hinf}
H_{\infty }\approx \sqrt{\frac{\Lambda_4 }{3}}\approx \omega
F_{\infty }
\end{equation}

\subsection{ \bf Linearized dynamical system:}

To solve the dynamical system (\ref{HF1}) and (\ref{HF2}), we
linearize the solution as in \cite{C1}:

\begin{equation}\label{H}
H=H_{\infty }+h(a)
\end{equation}

\begin{equation}\label{F}
F=F_{\infty }+f(a)
\end{equation}

where $\allowbreak $ $h(a)$ and $f(a)$ are linearized perturbation
functions to be determined later.

\bigskip Putting (\ref{H}) and (\ref{F}) into the field equations (\ref{HF1}) and
(\ref{HF2}) and neglecting higher terms in h(a) and f(a), one
obtains the following system:

\begin{equation}
\left(
\begin{array}{c}
h' \\
f'%
\end{array}%
\right) =-\frac{H_{\infty }}{\omega +1}\left(
\begin{array}{cc}
3\gamma \omega +4 & \left( \gamma -1\right) \omega \\
9\left( \gamma -1\right) & \left( 3\gamma +3\omega +1\right)%
\end{array}%
\right) \left(
\begin{array}{c}
h \\
f%
\end{array}%
\right) -\frac{k}{a^{2}}\left(
\begin{array}{c}
\frac{\left( 3\gamma \omega -2\omega +3\right) }{\left( 2\omega
+3\right) }
\\
\frac{3\left( 3\gamma -4\right) }{\left( 2\omega +3\right) }%
\end{array}%
\right)
\end{equation}
This system becomes
\begin{equation}
\left(
\begin{array}{c}
x' \\
y'%
\end{array}%
\right) =-\frac{1}{a}\left(
\begin{array}{cc}
\frac{3\omega +4}{\omega +1} & 0 \\
0 & \frac{\left( 3\gamma +3\gamma \omega +1\right) }{\omega +1}%
\end{array}%
\right) \left(
\begin{array}{c}
x \\
y%
\end{array}%
\right) +\left(
\begin{array}{c}
-\frac{3k}{a^{3}H_{\infty }\left( \omega +1\right) } \\
\frac{3k\left( 3\gamma -2\omega +3\gamma \omega -1\right)
}{a^{3}H_{\infty
}\left( \omega +1\right) \left( 2\omega +3\right) }%
\end{array}%
\right)
\end{equation}
with
\begin{equation}
h(a)=\frac{1}{3}\left( y\omega -x\right)
\end{equation}
and
\begin{equation}
f(a)=\left( x+y\right)
\end{equation}

Therefore, the linearized solutions (\ref{H}) and (\ref{F}), for
$\omega \rightharpoonup $ $\infty $, in the form as:
\begin{equation}\label{H2}
H=H_{\infty }-\frac{k}{2a_{0}^{2}H_{\infty }}\left(
\frac{a_{0}}{a}\right)
^{2}+H_{0}C_{1}\left( \frac{a_{0}}{a}\right) ^{3+\frac{1}{\omega }%
}+H_{0}C_{2}\left( \frac{a_{0}}{a}\right) ^{3\gamma
+\frac{1}{\omega }}
\end{equation}

\begin{equation}\label{F2}
F=F_{\infty }+H_{0}C_{1}\left( \frac{a_{0}}{a}\right)
^{3+\frac{1}{\omega }%
}+H_{0}C_{2}^{\prime }\left( \frac{a_{0}}{a}\right) ^{3\gamma
+\frac{1}{%
\omega }}
\end{equation}
where the subscript '0' indicates the present value. $C_{1}$,
$C_{2}$ and $C'_{2}=\frac{C_2}{\omega}$ are dimensionless
integration constants.\\
The equation (\ref{H2}) shows that the Hubble parameter varies
with the scale factor as in \cite{C1} but with an extra term whose
exponent depends on the $\gamma$-parameter.
\section{The cosmological parameters and dark energy:}

In this section we show that the Brans-Dicke theory is successful
in explaining the dark energy which we relate to the most
important cosmological parameters.

The Hubble parameter today, $H_{0}=\dot{a}/a(t_{0})$, is used to
estimate the order of magnitude for the present size and the age
of the universe, has the value:
\begin{equation}
H_{0}\equiv 100h\;km\;s^{-1}\;Mpc^{-1}
\end{equation}

In particular, we can define the individual ratios $\Omega
_{i}\equiv \rho _{i}/\rho _{c}$, for matter, radiation,
cosmological constant and even curvature, today. And from the
standard Friedmann equations we have \cite{B,K}:
\begin{equation}\label{H/H0}
\left( \frac{H}{H_{0}}\right) ^{2}=\Omega _{\Lambda }+\Omega
_{R}\left( \frac{a_{0}}{a}\right) ^{4}+\Omega _{M}\left(
\frac{a_{0}}{a}\right) ^{3}+\Omega _{k}\left(
\frac{a_{0}}{a}\right) ^{2}
\end{equation}

with
$$
\Omega _{\Lambda }=\frac{\Lambda_4 }{3H_{0}^{2}};  \ \ \ \ \ \ \ \
\ \ \ \Omega _{k}=-\frac{k}{a_{0}^{2}H_{0}^{2}};  \ \ \ \ \ \ \ \
\ \ \  \Omega _{M}=\frac{8\pi G\rho _{M}}{3H_{0}^{2}}; \ \ \ \ \
\ \ \ \ \ \ \Omega _{R}=\frac{8\pi G\rho _{R}}{3H_{0}^{2}}%
$$
Now  inserting the solution (\ref{H2}) in equation (\ref{H/H0})
one gets, in 4D space-time, the expression of the constants $C_1$
and $C_2$ by comparing respectively the expressions of $\Omega
_{i}$ in 4D with the BD ones in (\ref{H2}) for $\omega
\longrightarrow $ $\infty $.

First, we mention that all forms of matter/energy are possible.
However, we are interested in the $\gamma$-parameter of the
equation of state in order to recover the different exponents of
the equation (\ref{H/H0}). From (\ref{Hinf}) we obtain the
following result:

\begin{equation}
\left( \frac{H_{\infty }}{H_{0}}\right) ^{2}=\Omega _{\Lambda
}=\frac{%
\Lambda_4 }{3H_{0}^{2}} \;.
\end{equation}
The integration constants $C_1$ and $C_2$ for different values of
the $\gamma$-parameter are summarized as follows:

$$\begin{tabular}{|l|l|l|l|l|l|l|} \hline $\gamma $ & ${-1}/{3}$,
$0$, ${1}/{3}$ & ${1}/{2}$ & ${2}/{3}$ & $1$ & ${4}/{3}$ & $2$  \\
\hline $C_{1}$  & $\frac{\Omega _{M}}{2\sqrt{\Omega _{\Lambda }}}$
& $\frac{\Omega _{M}-C_2^2}{2\sqrt{\Omega _{\Lambda }}}$ &
$\frac{%
\Omega _{M}}{2\sqrt{\Omega _{\Lambda }}}$ & $\frac{\Omega
_{M}}{2\sqrt{%
\Omega _{\Lambda }}}-C_{2}$ & $\frac{\Omega _{M}}{2\sqrt{\Omega
_{\Lambda }}%
}$ & $\frac{\Omega _{M}}{2\sqrt{\Omega _{\Lambda }}}$ \\ \hline
$C_{2}$ &  $0$ & $\Omega _{M}-2C_1\sqrt{\Omega _{\Lambda }}$ & $0$
& $\frac{\Omega _{M}}{2\sqrt{\Omega
_{\Lambda }}}-C_{1}$ & $0$ & $\forall $  \\
\hline
\end{tabular}%
$$
we notice that in the cases $\gamma=\frac{1}{2}$ (extended
inflation) and $\gamma=1$ (dust universe) the integration
constants $C_1$ and $C_2$ depend on each other. The symbol
$\forall$ means that all values of $C_2$ are possible.

In the 3-brane, one obtains the same results by replacing the
preceding $\gamma $  by $\frac{\gamma}{2}$. Hence, $C_2$ and $C_1$
are related to each other for $\gamma=\frac{1}{4}$ and
$\gamma=\frac{1}{2}$.
\bigskip

To compare ours results with the experimental data, we use the
deceleration parameter $q_{0}={\Omega }_{R}+\frac{{1}}{2}{\Omega
}_{M}-{\Omega }_{\Lambda }$ \cite{Wein,B,K}. Neglecting ${\Omega
}_{R}$, one can parameterize the matter/energy content of
the universe with just two components: the matter, characterized by $%
{\Omega }_{M}$, and the vacuum energy by ${\Omega }%
_{\Lambda }$. \\
Except the cases $\gamma=\frac{1}{2},1$, where $C_1$ and $C_2$ are
not independent,
\begin{itemize}
\item  the line ${\Omega }_{\Lambda }=\frac{{\Omega }_{M}}{2}$,
separating accelerating from decelerating universe, corresponds
to:
\end{itemize}
\begin{equation}
C_{1}=\frac{\Omega _{M}}{2\sqrt{\Omega _{\Lambda }}}=\sqrt{\Omega
_{\Lambda }}
\end{equation}
$C_{1}< \sqrt{\Omega _{\Lambda }}$ corresponds to an
 accelerating universe, while
$C_{1}> \sqrt{\Omega _{\Lambda }}$ corresponds to a decelerating
universe.

\begin{itemize}
\item  the line $\Omega _{\Lambda }=1-{\Omega }_{M}$, separating
an open from a closed universes for a flat universe ($k=0$),
corresponds to
\end{itemize}
\begin{equation}
C_{1}=\frac{\Omega _{M}}{2\sqrt{\Omega _{\Lambda
}}}=\frac{1-\Omega _{\Lambda }}{\sqrt{\Omega _{\Lambda }}}
\end{equation}
$C_{1}< \frac{1-{\Omega }_{\Lambda }}{2\sqrt{\Omega _{\Lambda }}}$
corresponds to an open universe, while $C_{1}> \frac{1-{\Omega
}_{\Lambda }}{2\sqrt{\Omega _{\Lambda }}}$ corresponds to a closed
universe.\\

Except the cases where $C_1$ and $C_2$ are not independent,
experimental data give $C_{1}=\frac{\Omega _{M}}{2\sqrt{\Omega
_{\Lambda }}}\simeq 0.15$. So the $(\Omega _{M},\Omega _{\Lambda
})$ plane shows that we live in accelerating flat universe, since
$C_1<\sqrt{\Omega _{\Lambda }}$, which is in accordance with the
experimental data of Ia SNe \cite{Perl}. While in the 3-brane the
exception is for
$\gamma=\frac{1}{4}$ and $\gamma=\frac{1}{2}$. \\
The non vacuum dark energy, $\gamma=-\frac{1}{3}, \frac{1}{3},
\frac{1}{2}$ and $\frac{2}{3}$ in 4-dimension and
$\gamma=-\frac{1}{6}, \frac{1}{6}, \frac{1}{4}, \frac{1}{3},
\frac{1}{2}$ and $\frac{2}{3}$ in the 3-brane contribute strongly
to the behaviour of the universe as was seen before in the
dynamical of the universe, \\
we notice that the $\gamma$ spectrum is large in the 3-brane
compared to the one of the 4D space-time.\\
 We also  claim that the 3-brane allows a much wider
class of matter/energy than the 4D cosmology.
\section{Conclusions:}
In summary, we have shown that at high energy the dynamical
systems are the same in both cases with
$\gamma{(4-\rm{dimension})}$ replaced by
$2\gamma{(5-\rm{dimension})}$. However, at low energy the 5D BD
theory coincides with the 4D one. This means that in vacuum era
there is no difference between the 3-brane world and the
4-dimension universe. While in "stiff" matter ($\gamma=2$),
pressureless dust ($\gamma=1$) and radiation
($\gamma=\frac{4}{3}$) era in 4-dimension correspond to the dust,
extended inflation ($\gamma=\frac{1}{2}$) and
($\gamma=\frac{2}{3})$ era in 3-brane world. So in the 3-brane
universe, the era ($\gamma=1$), extended inflation era
($\gamma=\frac{1}{2},\frac{2}{3})$
 could be studied at high energy limit as stiff matter, pressureless and
 radiation era respectively in standard BD cosmology. From the values of
 the $\gamma$-parameter the non vacuum dark energy could be constituted by an exotic
 form of matter/energy or by a combined of ordinary form of matter/energy. This
 work opens a new perspective in this field of research, namely studying the extra
 dimensions of the universe as standard
 one with a modified equation of state. Finally it
 is also interesting to know what happens at
 intermediate energy limit, this will be a subject of a
 forthcoming paper.


\begin{thebibliography}{}
\bibitem{Green} M. B. Green, J. H. Schwarz and E. Witten,
Superstring Theory 1988 (Cambridge: Cambridge University Press).
\bibitem{Brans} C. Brans and R. H. Dicke, Phys. Rev. 124 (1961)
925.
\bibitem{Wein} S. Weinberg, Gravitation and Cosmology
(John Wiley \& Sons, San Francisco), 1972.
\bibitem{Rea} R. D. Reasenberg et al. Astrophys. J. 234,
(1979) L219.
\bibitem{EWein} E. J. Weinberg, Phys. Rev. D40 (1989) 3950.
\bibitem{La} D. La, P. J. Steinhardt and E. Bertschinger, Phys.
Lett. B231 (1989) 231.
\bibitem{Hol1} R. Holman, E. W. Kolb and Y. Wang, Phys. Rev.
Lett. 65 (1990) 17.
\bibitem{Hol2} R. Holman, E. W. Kolb, S. Vadas and Y. Wang, Phys.
Lett. B269 (1991) 252.
\bibitem{Bar} J. D. Barrow and K. Maeda, Nucl. Phys. B341 (1990)
294.
\bibitem{Stein} P. J. Steinhard and F. S. Accetta, Phys. Rev.
Lett. 64 (1990) 2740.
\bibitem{H1} P. Horava and E. Witten, Nucl. Phys. B460,
(1996) 506, hep-th/9510209.
\bibitem{H2} P. Horava and E. Witten, Nucl. Phys. B475,
(1996) 94, hep-th/9603142.
\bibitem{R1} L. Randall and R. Sundrum, Phys. Rev. Lett. 83,
(1999) 3370.
\bibitem{R2} L. Randall and R. Sundrum, Phys. Rev. Lett. 4690
(1999) 83.
\bibitem{Sch} T. Shiromizu, K. Maeda and M. Sasaki, Phys. Rev. D62 (2000) 024012.
\bibitem{Sp} D. N. Spergel \& al., Astrophys. J. Suppl. 148 (2003)
175.
\bibitem{Perl} S. J. Perllmutter et al. Astrophys. J. 517 (1999)
565, astro-ph/9812133.
\bibitem{Kn} R. A. Knop \& al., Astrophys.J. 598 (2003) 102.

\bibitem{K2} S. J. Kolitch, Annals Phys. 246 (1996) 121-132.
\bibitem{C1} M. Ar\'{i}k and M. C. \c{C}alik, gr-qc/0505035.
\bibitem{K1} S. J. Kolitch, Annals Phys. 241 (1995) 128-151.
\bibitem{W} J. P. Mimoso and D. Wands Phys. Rev. D51 (1995) 477-489.
\bibitem{Q} L. Qiang and al. Phys. Rev. D71 (2005) 061501.
\bibitem{B} J. Garcia-Bellido, Cosmology and Astrophysics,
astro-ph/0502139.
\bibitem{K} E. W. Kolb and M. S. Turner, The early universe, Addison Wesley
(1990).


\end{thebibliography}
\end{document}